\pdfoutput=1

\documentclass{tlp}
\usepackage{amsfonts,amsmath}
\usepackage{graphicx}
\usepackage{url}
\usepackage{verbatim} 
\usepackage{listings}
\usepackage{amsmath}
\usepackage{subfig}
\lstnewenvironment{codex}
    {\lstset{}%
      \csname lst@SetFirstLabel\endcsname}
    {\csname lst@SaveFirstLabel\endcsname}
\lstnewenvironment{code}
    {\lstset{}%
      \csname lst@SetFirstLabel\endcsname}
    {\csname lst@SaveFirstLabel\endcsname}
\newcommand{\BI}[0]{\begin{itemize}}
\newcommand{\EI}[0]{\end{itemize}}

\newcommand{\BE}[0]{\begin{enumerate}}
\newcommand{\EE}[0]{\end{enumerate}}

\newcommand{\BX}[0]{\begin{codex}}
\newcommand{\EX}[0]{\end{codex}}
\newcommand{\BV}[0]{\begin{verbatim}}
\newcommand{\EV}[0]{\end{verbatim}}

\def \bscale1 {0.25}
\def \bscale {0.25}

\newcommand{\mf}[1]{$#1$}

\renewcommand{\arraystretch}{1.5}

\newcommand{\progfont}[0]{\rm}
\newcommand{\K}[1]{{\progfont {#1}}} 

\newcommand{\Cell}[1]{{\framebox[4.5ex]{\strut #1}}}

\title[The BinProlog Experience]{
    The BinProlog Experience: Architecture and Implementation Choices
    for Continuation Passing Prolog and First-Class
    Logic Engines
}

\author[Paul Tarau]{
   Paul Tarau\\
   {Dept. of Computer Science and Engineering}\\
   {University of North Texas, Denton, Texas, USA}\\
   {\em E-mail: tarau@cs.unt.edu}
}

\pagerange{\pageref{firstpage}--\pageref{lastpage}}
\submitted{10th October 2009}
\revised{28st February 2010}
\accepted{2nd February 2011}

\begin{document}
\maketitle
\label{firstpage}

\begin{abstract}
We describe the {\em BinProlog} system's
compilation technology, runtime system and its extensions
supporting first-class Logic Engines while providing
a short history of its development, details of some of
its newer re-implementations as well as an
overview of the most important 
architectural choices involved in their
design.

With focus on its differences with
conventional WAM implementations, 
we explain key details of {\em BinProlog}'s compilation
technique, which replaces the WAM with
a simplified {\em continuation passing}
runtime system (the ``BinWAM"), based on a mapping of full 
Prolog to {\em binary logic programs}.
This is followed by a description of a 
{\em term compression} 
technique using a ``tag-on-data" representation.

Later derivatives, the Java-based {\em Jinni Prolog}
compiler and the recently developed {\em Lean Prolog}
system refine the {\em BinProlog}
architecture with {\em first-class Logic Engines},
made generic through the
use of an {\em Interactor} interface.
An overview of their applications
with focus  on the ability to express
at source level a wide variety
of Prolog built-ins and extensions,
covers these newer developments.

{\em To appear in Theory and Practice of Logic Programming (TPLP).}
\end{abstract}


\section{Introduction}

At the time when we started work on the BinProlog compiler, around
1991, WAM-based implementations \cite{WA83,kaci91:WAM}
had reached already a significant
level of maturity. The architectural changes occurring later
can be seen mostly as extensions for constraint programming
and runtime or compile-time optimizations.
 
BinProlog's design philosophy has been minimalistic from the very beginning.
In the spirit of Occam's razor, while developing
an implementation as an iterative process,
this meant not just trying to optimize
for speed and size, but also to actively look for
opportunities to refactor and simplify.

The guiding principle, at each stage, was seeking 
answers to questions like:
\begin{itemize}
\item what can be removed from the WAM
      without risking significant, program independent,
      performance losses?
\item what can be done to match, within small margins, performance gains
      resulting from new WAM optimizations (like  read/write stream separation,
      instruction unfolding, etc.) while minimizing 
      implementation complexity and code size?
\item can one get away with uniform data representations (e.g. no special tags for lists)
      instead of extensive specialization, without
      major impact on performance?
\item when designing new built-ins and extensions,
      can we use source-level transformations 
      rather than changes to the emulator?  
\end{itemize}
The first result of this design, BinProlog's {\em BinWAM} abstract machine 
has been originally implemented as a {\bf C} emulator based on a program
transformation introduced in \cite{Tarau90:PLILP}.

While describing it, we assume familiarity with the WAM and 
focus on the differences between the two abstract machines.
We refer to \cite{kaci91:WAM} for a tutorial description of the
WAM, including its instruction set, run-time areas and
compilation of unification and control structures.

The BinWAM replaces the WAM with a simplified continuation passing
logic engine \cite{Tarau91:JAP} based on a mapping of full 
Prolog to binary logic programs (binarization).
Its key assumption is that
as conventional WAM's environments are discarded in favor of a 
heap-only run-time system,
heap garbage collection and efficient term representation become instrumental
as means to ensure ability to run large classes of Prolog programs. 

The second architectural novelty, present to some extent in
the original BinProlog and a key element of its newer
Java-based derivatives Jinni Prolog \cite{tarau:shaker,j2k_ug} and Lean Prolog
(still under development) is the use of Interactors
(and first-class Logic Engines, in particular) as a uniform
mechanism for the source-level specification (and often actual
implementation) of key built-ins and language extensions
\cite{tarau:cl2000,iclp08:inter,padl09inter}.

We first explore
various aspects of the compilation
process and the runtime system. Next we discuss
source-level specifications of key built-ins
and extensions using first-class Logic Engines.


Sections \ref{bin} and \ref{bincomp} provide an overview BinProlog's key
source-to-source transformation ({\em binarization})
and its use in compilation. 

Section
\ref{data} introduces BinProlog's unusual
``tag-on-data'' term representation (\ref{tag}) and
studies its impact on term compression (\ref{compress}).

Section \ref{interp} discusses
optimizations of the runtime
system like instruction compression 
and the implicit handling of read-write modes. 

Section \ref{interactors} introduces 
Logic Engines seen as implementations of
a generic Interactor interface
and describes their basic operations.

Section \ref{srcext} applies Interactors
to implement, at source level, some 
key Prolog built-ins, exceptions (\ref{exc})
and higher order constructs (\ref{higher}).

Section \ref{kernelext} applies
the logic engine API to specify
Prolog extensions ranging
from dynamic database operations (\ref{db})
and backtracking if-then-else (\ref{ifany})
to predicates comparing alternative answers \ref{altanswers}
and mechanisms to encapsulate
infinite streams \ref{infinite}.


Section \ref{history} gives a short historical account
of BinProlog and its derivatives.

Section
\ref{related} discusses related
work and Section \ref{concl}
concludes the paper.

\section{The binarization transformation} \label{bin}

We start by reviewing the program transformation
that allows compilation of logic programs towards a
{\em simplified WAM specialized for the execution of binary 
programs} (called BinWAM from now on).
Binary programs consist
of facts and binary clauses that have only one atom in the body 
(except for some inline ``built-in" operations like arithmetics)
and therefore they need no ``return" after a call.
A transformation introduced in \cite{Tarau90:PLILP} allows the
emulation of logic programs with operationally equivalent
binary programs.

Before defining the {\em binarization} transformation, we describe two
auxiliary transformations, commonly used by Prolog compilers.

The first transformation converts facts into rules by  giving
them the atom {\tt true} as body. E.g., the fact {\tt p} is
transformed into the rule {\tt p :- true}.

The second transformation eliminates {\em metavariables} (i.e. variables 
representing Prolog goals only known at run-time), by wrapping them 
in a {\tt call/1} predicate, 
e.g., a clause like {\tt and(X,Y):-X,Y} is transformed into {\tt 
and(X,Y) :- call(X),call(Y).}

The {\em binarization transformation} (first described in \cite{Tarau90:PLILP})
adds continuations
as  the last argument of predicates in a way that  preserves
first argument indexing.

Let   $P$ be  a definite  program  and \mf{Cont}  a  new
variable. Let  \mf{T} and \mf{E=p(T_1,...,T_n)} be  two 
expressions (i.e. atoms or terms). We  denote by
\mf{\psi(E,T)} the expression \mf{p(T_1,...,T_n,T)}. 
Starting with the clause

\verb~(C)~  \qquad \mf{A :- B_1,B_2,...,B_n.}

\noindent we construct the clause

\verb~(C')~  \qquad \mf{\psi(A,Cont) :-
 \psi(B_1,\psi(B_2,...,\psi(B_n,Cont))).}
                         
\noindent
The set $P'$ of all clauses \verb~C'~ obtained from the clauses of $P$ is called
the binarization of $P$. 

The following example shows the result of this
transformation on the well-known ``naive reverse" program:

\begin{verbatim}
   app([],Ys,Ys,Cont):-true(Cont).
   app([A|Xs],Ys,[A|Zs],Cont):-app(Xs,Ys,Zs,Cont).
                                  
   nrev([],[],Cont):-true(Cont).
   nrev([X|Xs],Zs,Cont):-nrev(Xs,Ys,app(Ys,[X],Zs,Cont)).
\end{verbatim}
Note that {\tt true(Cont)} can be seen as a (specialized version) of Prolog's {\tt call/1} that executes the goals stacked in the continuation variable {\tt Cont}. Its semantics is expressed by the following clauses
\begin{code}
   true(app(X,Y,Z,Cont)):-app(X,Y,Z,Cont).
   true(nrev(X,Y,Cont)):-nrev(X,Y,Cont).
   true(true).
\end{code}
which, together with the code for {\tt nrev} and {\tt app} run the binarized query
\begin{codex}
?- nrev([1,2,3],R,true).
\end{codex}
in any Prolog, directly, returning {\tt R=[3,2,1].}

Prolog's inference rule (called LD-resolution) 
executes goals in the
body of a clause left-to-right in a depth first order.
LD-resolution describes Prolog's operational 
semantics more accurately than order-independent
SLD-resolution \cite{LL87}. 
The binarization transformation preserves
a strong operational equivalence with the
original program with respect to the LD-resolution rule which
is {\em reified} in the syntactical structure of the resulting program,
where the order of the goals in the body becomes hardwired in the
representation \cite{pt93b}. This means that each resolution step of 
an LD-derivation on a definite program $P$
can be mapped to an LD-resolution step of the binarized program $P'$.
{\em
More precisely, let $G$ be an atomic goal and \mf{G'=\psi(G,true)}. Then, the answers computed
using LD-resolution obtained by
querying $P$ with $G$ 
are the same as those obtained by querying $P'$ with $G'$.
} Note also that the concepts of SLD- and LD-resolution overlap in the
case of binary programs.

\section{Binarization based compilation and runtime system} \label{bincomp}

BinProlog's BinWAM virtual machine specializes the WAM to binary clauses
and therefore it drops WAM's environments. Alternatively, assuming a two stack WAM implementation
the BinWAM can be seen as an OR-stack-only WAM. 
Independently, its simplifications of the indexing mechanism and
a different ``tag-on-data" representation are the most important differences with conventional WAM
implementations. The latter also brings opportunities for a more compact heap representation
that is discussed in section \ref{data}. 

Note also that continuations become explicit in the binary version of the program.
We refer to \cite{td94:LOPSTR} for a technique to access and manipulate them
by modifying BinProlog's binarization preprocessor. This results in the ability to
express constructs like a backtracking sensitive variant of catch/throw at source level.
We focus in this section only on their uses in BinProlog's compiler and 
runtime system.

\subsection{Metacalls as built-ins} \label{meta}

The first step of our compilation process simply wraps metavariables inside 
a predicate \verb~call/1~, and adds {\tt true/0} as a body for facts, as most Prolog compilers do.
The binarization transformation then adds a continuation as last 
arguments of each predicate and a new predicate \verb~true/1~ to deal with 
unit clauses. During this step, the arity of all predicates
increases by {\tt 1} so that, for instance, \verb~call/1~ becomes \verb~call/2~.

Although we can add the special clause {\tt true(true)}, 
and for each functor {\tt f} occurring in the program, clauses like
\begin{codex}
true(f(...,Cont)):-f(...,Cont).
call(f(...),Cont):-f(...,Cont).
\end{codex}
as an implementation of {\tt true/1} and {\tt call/2},
in practice it is simpler and more 
efficient to treat them as built-ins \cite{Tarau91:JAP}.

The built-in corresponding to {\tt true/1}
looks up the address of the predicate associated to
{\tt f(...,Cont)}
and throws an exception if no such predicate
is found.
The built-in corresponding to {\tt call/2}
adds the extra argument {\tt Cont} to {\tt f(...)},
looks up wether a predicate definition
is associated to {\tt f(...,Cont)}
and throws an exception if no definition is found.
In both cases, when predicate definitions are found,
the BinWAM fills up the argument registers
and proceeds with the execution of the
code of those predicates.

Note that the predicate look-ups are implemented efficiently by using
hashing on a {\tt <symbol, arity>} pair stored
in one machine word. Moreover, they happen
relatively infrequently.
For the case of {\tt call/2}-induced look-ups, as in ordinary Prolog
compilation, they are generated only
when metavariables are used. As calls to {\tt true/1} only
happen when execution reaches a ``fact" in the original program, they also have 
a relatively little impact on performance, for typical 
recursion intensive programs.

\subsection{Inline compilation of built-ins} \label{built-ins}

Demoen and Mari\"{e}n pointed out in \cite{Demoen91:RU} that a more 
implementation oriented view of binary programs can be very useful: a binary 
program is simply one that does not need an environment in the WAM. This 
view leads to inline code generation (rather than binarization)
for built-ins occurring {\em immediately after the head}.
For instance something like
\begin{codex}
a(X):-X>1,b(X),c(X).
\end{codex}
is handled as:
\begin{codex}
a(X,Cont) :- inline_code_for(X>1),b(X,c(X,Cont)).
\end{codex}
rather than
\begin{codex}
a(X,Cont) :- '>'(X,1,b(X,c(X,Cont))).
\end{codex}
Inline expansion of built-ins
contributes significantly to BinProlog's speed and
supports the equivalent of WAM's last call optimization for frequently occurring
linear recursive predicates containing such built-ins,
as unnecessary construction of continuation 
terms on the heap is avoided for them.

\subsection{Handling CUT}
Like in the WAM, a special register {\tt cutB} contains
the choice point address up to where
choice points need to be popped off
the stack on backtracking.
In clauses like
\begin{code}
a(X):-X>1,!,b(X),c(X).
\end{code}
{\tt CUT} can be handled inline (by a special instruction {\tt PUSH\_CUT}, generated when the compiler recognizes this
case), 
after the built-in {\tt X>1}, by
trimming the choice point stack right away. On the other hand,
in clauses like
\begin{code}
a(X):-X>1,b(X),!,c(X).
\end{code}
a pair of instructions {\tt PUT\_CUT} and {\tt GET\_CUT} is needed.
During the BinWAM's term creation, {\tt PUT\_CUT} saves 
to the heap the register {\tt cutB}. This value of {\tt cutB} is 
used by {\tt GET\_CUT} when
the execution of the instruction sequence reaches it,
to trim the choice point stack to the appropriate level.

\subsection{Term construction in the absence of the AND-stack} \label{early}

The most important simplification in the BinWAM in comparison with the standard WAM is the absence of an AND-stack. 
Clearly, this is made possible by the fact that each binary clause has (at most) one goal in the body.

In procedural and call-by-value functional languages 
featuring only deterministic
calls it was a typical implementation choice
to avoid repeated structure creation by using
environment stacks containing only the variable bindings. The
WAM \cite{WA83} 
follows this model based on the assumption that most logic programs
are deterministic.

This is one of the key points where the execution models between 
the WAM and BinWAM differ.
A careful analysis suggests that the choice between
\begin{itemize}
  \item the standard WAM's late and repeated construction with variables of each goal in the body pushed on the AND stack
  \item the BinWAM's eager early construction on the heap (once) and reuse (by possibly trailing/untrailing variables)
\end{itemize}
favors different programming styles, with ``AND-intensive", deterministic, 
possibly tail-recursive programs favoring the WAM while ``OR-intensive", nondeterministic programs
reusing structures through trailing/untrailing favoring the BinWAM.
The following example illustrates the difference between the two execution models. In the clause
\begin{code}
  p(X) :- q(X,Y), r(f(X,Y)).
\end{code}
binarized as
\begin{code}
  p(X,C) :- q(X,Y,r(f(X,Y),C)).
\end{code}
the term {\tt f(X,Y)} is created on the heap by the WAM
as many times as the number of solutions of the predicate 
{\tt q}. On the other hand,  the BinWAM creates it only once and reuses it
by undoing the bindings of variables {\tt X} and {\tt Y} (possibly trailed).
This means that if {\tt q} fails, the BinWAM's ``speculative" term creation work is wasted.
And it also means that if {\tt q} is nondeterministic and has a large
number of solutions, then the WAM's repeated term creation leads
to a less efficient execution model.

\subsection{A minimalistic BinWAM instruction set}
A minimalistic BinWAM instruction set (as shown for two simple {\bf C} and {\tt Java} implementations
at \url{http://www.binnetcorp.com/OpenCode/free_prolog.html})
consists of the following subset of the WAM: {\tt GET\_STRUCTURE, UNIFY\_VARIABLE,
UNIFY\_VALUE, EXECUTE, PROCEED, TRY\_ME\_ELSE, RETRY\_ME\_ELSE, TRUST\_ME},
as well as the following instructions, that the reader will recognize as
mild variations of their counterparts in the ``vanilla" WAM instruction set \cite{kaci91:WAM}.
\begin{itemize}
\item MOVE\_REGISTER (simple register-to-register move)
\item NONDET (sets up choice-point creation when needed)
\item SWITCH (simple first-argument indexing)
\item PUSH\_CUT, PUT\_CUT, GET\_CUT (cut handling instructions for binary programs,
along the lines of \cite{Demoen91:RU})
\end{itemize}
Note that specializations for CONSTANTs, LISTs as well as WRITE-mode variants of the GET and UNIFY instructions can be added as obvious optimizations.

\subsection{The OR-stack} \label{runtime}

A simplified {\em OR-stack} having the layout shown 
in Figure ~\ref{stack} is used only for (1-level)
{\em choice point creation} in nondeterministic predicates.
No link pointers between frames are needed as the length of
the frames can be derived from the arity of the predicate.

\begin{figure}
\centering
\begin{tabular}{||ll||}
P $\Rightarrow$ & next clause address \\ 
H $\Rightarrow$ & saved top of the heap \\ 
TR $\Rightarrow$ & saved top of the trail \\ 
$A_{N+1}$ $\Rightarrow$ & continuation argument register \\
$A_{N}$ $\Rightarrow$ & saved argument register N \\ 
... & ... \\
$A_{1}$ $\Rightarrow$ & saved argument register 1 \\ 
\end{tabular}
\medskip
\caption{A frame on BinProlog's OR-stack. \label{stack}}
\end{figure}

Given that
variables kept on the local stack in conventional WAM are now located
on the heap, the heap consumption of the program increases.
It has been shown that,
in some special cases, partial evaluation at source level can deal with 
the problem \cite{DBLP:conf/lopstr/Demoen92,Neum92} but
as a more practical solution, the impact of heap consumption has been 
alleviated in BinProlog by the use of an
efficient copying garbage collector \cite{Demoen96:GC}.


\subsection{A simplified clause selection and indexing mechanism} \label{indexing}

As the compiler works on a clause-by-clause basis, it is the responsibility of the 
loader (that is part of the runtime system) to index clauses and link the code. 
The runtime system uses a global $<key_1,key_2> \rightarrow value$ hash table seen 
as an abstract {\em multipurpose dictionary}.
This dictionary provides the following services:
\begin{itemize}
 \item indexing compiled code, with $key_1$ as the functor of the predicate and $key_2$
as the functor of the first argument
\item implementing
multiple dynamic databases, with $key_1$ as the name of the database
and $key_2$ the functor of a dynamic predicate
\item  supporting a user-level storage area (called ``blackboard") containing global terms
indexed by two keys
\end{itemize}
A one byte mark-field in the table
is used to distinguish between {\em load-time} use when the {\em kernel} (including
built-ins written in Prolog and the compiler itself) is loaded,
and {\em run-time} use (when user programs are compiled and loaded) to protect against
modifications to the kernel and 
for fast clean-up. 
Sharing of the global multipurpose dictionary, although 
somewhat slower than the small $key \rightarrow value$ hashing tables 
injected into the code-space of the standard WAM, 
keeps the implementation as simple as possible.
Also, with data areas of fixed size (as in the original BinProlog implementation), one big
dictionary provides overall better use of the available memory
by sharing the hashing table for different purposes.

Predicates are classified as {\em single-clause}, {\em deterministic} 
and {\em nondeterministic}. Only predicates having {\em all first-argument functors 
distinct} are detected as deterministic and indexed.

In contrast to the WAM's fairly elaborate indexing mechanism,
indexing of deterministic predicates in the BinWAM is done by a unique SWITCH
instruction. 

If the first argument dereferences to a non-variable, SWITCH 
either fails or finds the 1-word address of the unique matching clause in 
the global hash-table, using the {\em predicate} and 
the {\em functor of the first argument} as a 2-word key. 
Note that the basic difference with the WAM is the 
absence of intensive tag analysis. This is related also to our different 
low-level data-representation that we  discuss in section \ref{data}.

A specialized JUMP-IF instruction deals with the frequent case 
of 2 clause deterministic predicates. To reduce the interpretation 
overhead, SWITCH and JUMP\_IF are combined with the 
preceding EXECUTE and the following GET\_STRUCTURE 
or GET\_CONSTANT instruction, giving EXEC\_SWITCH 
and EXEC\_JUMP\_IF. This not only avoids dereferencing 
the first argument twice, but also reduces unnecessary branching logic
that breaks the processor's pipeline. 

Note also that simplification of the indexing
mechanism, in combination with 
smaller and unlinked choice points,
helps making backtracking sometimes 
faster in the BinWAM than in conventional WAMs, as in the case
of simple (but frequent!) predicates having only a few clauses.

However, as mentioned in subsection \ref{early}, backtracking performance
in the BinWAM also benefits from sharing
structures occurring in the body of a clause in the OR-subtree it generates,
instead of repeated creation as in conventional WAM. This
property of binarized programs (see example in subsection \ref{early}),
was first pointed out in \cite{Demoen91:RU} as
the typical case when binarized variants are faster than the original
programs.

Our original assumption when simplifying the WAM's
indexing instructions was that, for predicates having a more general 
distribution of first-arguments, a source-to-source transformation,
grouping similar arguments into new predicates,
can be used. 

Later in time, while noticing that often
well written Prolog code tends to be either
``database type" (requiring multiple argument indexing)
or ``recursion intensive" (with small predicates
having a few clauses, fitting well this simplified first
argument indexing mechanism) it became clear that
it makes sense to handle these two problems separately.
As a result, we have kept this simplified indexing scheme
(for ``recursion intensive" compiled code) 
unchanged through the evolution of BinProlog and its derivatives.
On the other hand, our newest implementation, {\em Lean
Prolog} handles ``database type" dynamic code efficiently using
a very general multi-argument indexing mechanism.

\subsection{Binarization: some infelicities}
We have seen that binarization has helped building a simplified abstract machine
that provides good performance with help from a few low level optimizations. However,
there are some ``infelicities" that one has to face, somewhat similar to
what any program transformation mechanism induces at runtime - and DCG grammars
come to one's mind in the Prolog world.

For instance, the execution order in the body is reified at
compile time into a fixed structure. This means that things like
dynamic reordering of the goal in a clause body or 
AND-parallel execution mechanisms become trickier.
Also, inline compilation of constructs like if-then-else becomes
more difficult - although one can argue that using
a source-level technique, when available (e.g.
by creating small new predicates) is an acceptable implementation
in this case.

\section{Data representation} \label{data}

We review here an unconventional data representation choice that turned
out to also provide a surprising term-compression mechanism, that can
be seen as a generalization of ``CDR-coding'' \cite{Clark:1977:ESL:359423.359427}
used in LISP/Scheme systems.

\subsection{Tag-on-pointer versus tag-on-data} \label{tag}

When describing the data in a cell with a tag we have basically 2 possibilities. 
We can put a tag in the pointer to the data or in the data cell itself.

The first possibility, probably most popular among WAM implementors, allows one to 
check the tag before deciding {\em if} and {\em how} it has to be processed. 
We choose the second possibility as in the presence of indexing, unifications 
are more often intended to succeed propagating bindings, rather than being 
used as a clause selection mechanism. This also justifies why we have not
 implemented the WAM's traditional \verb~SWITCH_ON_TAG~ instruction.

We found it very convenient to precompute a functor in the code-space as a 
word of the form \verb~<arity,symbol-number,tag>~ \footnote{This technique is
also used in various other Prologs e.g. SICStus, Ciao.} and then simply compare 
it with objects on the heap or in registers. In contrast, in a conventional
WAM, one compares the tags, 
finding out that they are almost always the same, then compares the 
functor-names and finally 
compares the arities - an unnecessary but costly if-logic. This is avoided
with our {\em tag-on-data} representation, while also 
consuming as few tag bits as possible. Only 2 bits are used in BinProlog for tagging
{\em variables}, {\em integers} and {\em functors/atoms}\footnote{This representation
limits arity and available symbol numbers - a problem that,
went away with the newer
64-bit versions of BinProlog.}.
With this representation, a functor fits completely in one word.

As an interesting consequence, as we have found out later, when implementing
a symbol garbage collector for a derivative of BinProlog, the ``tag-on-data''
representation makes scanning the heap for symbols (and updating them in place)
a trivial operation. 

\subsection{Term compression} \label{compress}
\begin{figure}
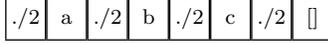

{\footnotesize
\Cell{./2}\Cell{a}\Cell{./2}%
\Cell{b}\Cell{./2}\Cell{c}%
\Cell{./2}\Cell{[]}
}
\medskip
\caption{Compressed list representation of [a,b,c] \label{list}}
\end{figure}

If a term has a last argument containing a functor, with our tag-on-data representation 
we can avoid the extra pointer from the last argument to
the functor cell and simply make them collapse. Obviously the unification algorithm 
must take care of this case, but the space savings are important, especially in the 
case of lists which become contiguous vectors with their N-th element directly 
addressable at offset \verb~2*sizeof(term)*N+1~ bytes from the beginning of 
the list, as shown in Figure \ref{list}.

\begin{figure}
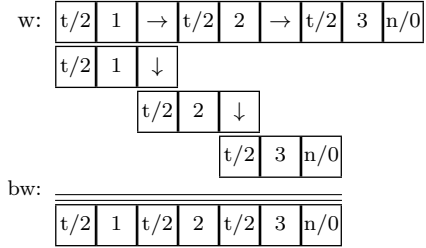

\begin{quote}
\begin{enumerate}
\item[w:]{\footnotesize
\Cell{t/2}\Cell{1}\Cell{\K{$\rightarrow$}}%
\Cell{t/2}\Cell{2}\Cell{\K{$\rightarrow$}}%
\Cell{t/2}\Cell{3}\Cell{n/0}
}

\item[]{\footnotesize%
\newlength{\leerzelle}%
\settowidth{\leerzelle}{\Cell{t/2}}%
\newcommand{\Empt}[1]{\makebox[\leerzelle]{\strut}}

\Cell{t/2}\Cell{1}\Cell{$\downarrow$}

\Empt{t/2}\Empt{1}\Cell{t/2}\Cell{2}\Cell{$\downarrow$}

\Empt{t/2}\Empt{1}\Empt{t/2}\Empt{2}\Cell{t/2}\Cell{3}\Cell{n/0}
\item[bw:]
\settowidth{\leerzelle}{\Empt{t/2}\Empt{1}\Empt{t/2}\Empt{2}\Cell{t/2}\Cell{3}\Cell{n/0}}

\makebox[\leerzelle]{\hrulefill} \\[-2.4mm]
\makebox[\leerzelle]{\hrulefill}

\Cell{t/2}\Cell{1}%
\Cell{t/2}\Cell{2}%
\Cell{t/2}\Cell{3}\Cell{n/0}
}
\end{enumerate}
\end{quote}
\medskip
\caption{Term compression. w: WAM, bw: BinWAM. \label{tc}}
\end{figure}

The effect of this {\em last argument overlapping} on
{\tt t(1,t(2,t(3,n)))} is represented in Figure ~\ref{tc}.

This representation also reduces the space consumption for lists and other 
``chained functors'' to values similar or better than in the case of conventional WAMs. 
We refer to \cite{Tarau93:comp} for the details of the term-compression related 
optimizations of BinProlog.

\section{Optimizing the run-time system} \label{interp}
We give here an overview of the optimizations of the runtime system. 
Most of them are, at this point in time, ``folklore" and
shared with various other WAM-based Prolog implementations.

\paragraph{Instruction compression} \label{instr}
It happens very often that a sequence of consecutive instructions share some 
WAM state information \cite{Nassen:2001:IMS:773184.773191}. 
For example, two consecutive unify instructions have the 
{\em same mode} as they correspond to arguments of the same structure. Moreover, due 
to our very simple instruction set, some instructions have only a few possible other
instructions that can follow them. For example, after an EXECUTE instruction, 
we can have a single, a deterministic or a nondeterministic clause.
 It makes sense to specialize the EXECUTE instruction with respect to what 
 has to be done in each case. This gives, in the case of calls to deterministic 
 predicates the instructions EXEC\_SWITCH and EXEC\_JUMP\_IF as mentioned 
 in the section on indexing. On the other hand, some instructions are simply so 
 small that just dispatching them can cost more than actually performing the 
 associated WAM-step.

This in itself is a reason to compress two or more instructions taking less than 
a word in one instruction. 
This optimization has been part of WAM-based Prolog systems like
Quintus, SICStus, Ciao as well.
Also having a small initial instruction set reduces the number of combined
instructions needed to cover all cases.
For example, by compressing our UNIFY instructions 
and their WRITE-mode specializations, we get the new instructions:
 
\begin{codex}
UNIFY_VARIABLE_VARIABLE
WRITE_VARIABLE_VARIABLE
...
\end{codex}

This gives, in the case of the binarized version of the recursive clause of 
append/3, the following code:

\begin{codex}
append([A|Xs],Ys,[A|Zs],Cont):-append(Xs,Ys,Zs,Cont).

TRUST_ME_ELSE */4,     % keeps also the arity = 4
GET_STRUCTURE X1, ./2
UNIFY_VARIABLE_VARIABLE X5, A1
GET_STRUCTURE X3, ./2
UNIFY_VALUE_VARIABLE X5, A3
EXEC_JUMP_IF  append/4 % actually the address of append/4
\end{codex}

The choice of candidates for instruction compression was based on
low level profiling (instruction frequencies) and possibility 
of sharing of common work by two successive instructions and frequencies
of functors with various arities.

BinProlog  also integrates the preceding GET\_STRUCTURE instruction 
into the double UNIFY instructions and the preceding PUT\_STRUCTURE 
into the double WRITE instructions. This gives another 16 instructions but it 
covers a large majority of uses of GET\_STRUCTURE and PUT\_STRUCTURE. 
\begin{codex}
GET_UNIFY_VARIABLE_VARIABLE
...
PUT_WRITE_VARIABLE_VALUE
....
\end{codex}
Reducing interpretation overhead on those critical, high frequency instructions 
definitely contributes to the speed of our emulator. As a consequence, in the 
frequent case of structures of arity=2 (lists included), 
mode-related IF-logic is completely eliminated, with up to 50\verb~%~
speed improvements for simple predicates like {\tt append/3}. 

The following example shows the
effect of this transformation:

\begin{codex}
a(X,Z):-b(X,Y),c(Y,Z). =>binary form=> a(X,Z,C):-b(X,Y,c(Y,Z,C)).

BinProlog BinWAM code, without compression

a/3:
PUT_STRUCTURE        X4<-c/3
WRITE_VARIABLE       X5
WRITE_VALUE          X2
WRITE_VALUE          X3
MOVE_REG             X2<-X5
MOVE_REG             X3<-X4
EXECUTE              b/3

BinProlog BinWAM code, with instruction compression
                       
PUT_WRITE_VARIABLE_VALUE  X4<-c/3, X5,X2
WRITE_VALUE               X3
MOVE_REGx2                X2<-X5, X3<-X4
EXECUTE                   b/3
\end{codex}

Note that instruction compression is usually applied inside a procedure. 
As BinProlog has a unique primitive EXECUTE instruction instead of 
standard WAM's CALL, ALLOCATE, DEALLOCATE, EXECUTE, PROCEED 
we can afford to do instruction compression across procedure boundaries 
with very little increase in code size due to relatively few different ways to 
combine control instructions.
Inter-procedural instruction compression can be seen as a 
kind of ``hand-crafted" {\em partial evaluation} at implementation language
level, intended to optimize the main loop of the WAM-emulator. 
It has the same effect as {\em partial evaluation} at source level which also 
eliminates procedure calls.
At the global level, knowledge about possible continuations can also 
remove the run-time effort of address look-up for meta-variables in predicate positions,
and of useless 
trailing and dereferencing.

\paragraph{(Most of) the benefits of two-stream compilation for free} \label{streams}
Let us point out here 
that in the case of GET\_*\_* instructions we have the benefits of
separate READ and WRITE streams
(for instance, avoidance of mode checking)
on some high frequency
instructions without actually incurring the compilation complexity
and emulation overhead in generating them.
As terms of depth 1 and functors of low arity
dominate statistically Prolog programs,
we can see that our instruction compression scheme actually
behaves as if two separate instruction streams were
present, most of the time!

\section{Logic Engines as Interactors} \label{interactors}

We now turn the page to a historically later
architectural feature of BinProlog and its newer derivatives. 
While orthogonal to the BinWAM
architecture, it shares the same philosophy: 
proceed with a fundamental system simplification on
purely {\em esthetic grounds},
independently of short term 
performance concerns, and hope
that overall elegance will
provide performance improvements
for free, later\footnote{
Even in cases when
such hopes do not materialize,
indirect consequences of such
architectural simplifications often
lower software risks and bring increased system
reliability while keeping implementation
effort under control.
}.

BinProlog's Java-based reimplementation, {\em Jinni} has
been mainly used in various 
applications \cite{T98:jelia,tarau:paam99,tarau:shaker,iclp04:jinni}
as an {\em intelligent agent infrastructure},
by taking advantage of Prolog's knowledge
processing capabilities in combination with a simple
and easily extensible runtime kernel supporting
a flexible reflexion mechanism \cite{padl_java}.
Naturally, this has suggested to investigate whether some
basic agent-oriented language design ideas can be 
used for a refactoring of pure Prolog's 
interaction with the external world.

Agent programming constructs have influenced design patterns 
at ``macro level'', ranging from interactive Web services to 
mixed initiative computer human interaction. {\em Performatives} 
in Agent communication languages 
\cite{fipa2:97} 
have made these constructs reflect explicitly the intentionality, 
as well as the negotiation process involved in agent interactions. 
At the same time, it has been a long tradition of logic programming languages
\cite{12071,Lusk93applicationsof}
to use multiple Logic Engines for supporting
concurrent execution.

In this context, the Jinni Prolog agent programming 
framework  \cite{ciclops:jinni} and the recent versions of the BinProlog 
system \cite{bp7advanced} have been centered 
around logic engine constructs providing an API that supports 
reentrant instances of the language processor. 
This has naturally led to a view of Logic Engines as instances 
of a generalized family of iterators called {\em Fluents} 
\cite{tarau:cl2000}, that have allowed the separation of the 
first-class language interpreters from the multi-threading 
mechanism, while providing, at the same time, 
a very concise source-level 
reconstruction of Prolog's built-ins. Later we have extended
the original {\em Fluents} with a few new operations
\cite{padl09inter} supporting bi-directional, mixed-initiative
exchanges between engines.

The resulting language constructs, that we have called {\em Interactors}, 
express coroutining, metaprogramming and interoperation with 
stateful objects and external services.
They complement pure Horn Clause Prolog with a significant
boost in expressiveness, to the point where they
allow emulating at source level virtually all Prolog
built-ins, including dynamic database operations.

In a wider programming language implementation
context, a {\tt yield} statement supports basic coroutining
in newer object oriented languages  like 
{\tt Ruby} 
\verb~C#~ 
and {\tt Python} 
but it goes back as far as \cite{coroutinesconway1963} 
and the {\em Coroutine Iterators} introduced in older 
languages like CLU \cite{DBLP:books/sp/Liskov81}.

\subsection{Logic Engines as answer generators} \label{logeng}

Our {\em Interactor API}, a unified interface to various stateful
objects interacting with Prolog processors, has evolved progressively into
a practical Prolog implementation framework starting with 
\cite{tarau:cl2000} and continued
with \cite{iclp08:inter} and \cite{padl09inter}.
We summarize it here while instantiating the more 
general framework to focus on
interoperation of Logic Engines.
We refer to \cite{padl09inter} for the details of an emulation
in terms of Horn Clause Logic of various engine operations.

An {\em Engine} is simply a 
language processor reflected through an API that 
allows its computations to be controlled 
interactively from another Engine
very much the same way a programmer controls Prolog's interactive 
toplevel loop: launch a new goal, ask for a new answer, 
interpret it, react to it. 
A {\em Logic Engine} is an Engine 
running a Horn Clause Interpreter 
with LD-resolution \cite{Tarau93:CONS} 
on a given clause database, together with a set of 
built-in operations.
The command
\begin{codex}
new_engine(AnswerPattern,Goal,Interactor)
\end{codex}
{\noindent  creates} a new Horn Clause solver, uniquely identified
by {\tt Interactor}, which shares code
with the currently running program and is initialized
with {\tt Goal} as a starting point. 
{\tt AnswerPattern} is a term, usually
a list of variables occurring in {\tt Goal},
of which answers
returned by the engine will be instances.
Note however that {\tt new\_engine/3} acts like a typical
constructor, no computations are performed at this
point, except for allocating data areas.

In our newer implementations, with all data
areas dynamic, engines are lightweight and
engine creation is fast and memory efficient\footnote{
The additional operation {\tt 
load\_engine(Interactor,AnswerPattern,Goal)} that
clears data areas and initializes an engine with {\tt AnswerPattern,Goal}
has also been available as a further optimization, by providing a
mechanism to reuse an existing engine.
} to the point where using them as building blocks
for a significant number of built-ins and 
various language constructs is not always
prohibitive in terms of performance.

\subsection{Iterating over computed answers}

Note that our Logic Engines are seen, in
an object oriented-style, as implementing the {\em interface} 
{\tt Interactor}. This supports a {\em uniform}
interaction mechanism with a variety of objects ranging from Logic Engines to
file/socket streams and iterators over external data structures.

The {\tt get/2} operation is used to retrieve successive 
answers generated by an Interactor, on demand.
It is also responsible for actually 
triggering computations in the engine. The query
\begin{codex}
get(Interactor,AnswerInstance)
\end{codex}
\noindent tries to harvest the answer computed from {\tt Goal}, 
as an instance of {\tt AnswerPattern}. If an answer
is found, it is returned as {\tt the(AnswerInstance)}, 
otherwise the atom {\tt no} 
is returned. 
As in the case of the {\tt Maybe} Monad in Haskell, 
returning distinct functors in the case of success and 
failure, allows
further case analysis in a pure Horn Clause style, 
without needing Prolog's CUT or if-then-else operation. 

Note that bindings are not propagated to the original {\tt Goal}
or {\tt AnswerPattern} when {\tt get/2} retrieves an answer,
i.e. {\tt AnswerInstance} is obtained by first standardizing apart
(renaming) the variables in {\tt Goal} and {\tt AnswerPattern}, and then
backtracking over its alternative answers in a separate Prolog
interpreter. Therefore, backtracking in the caller does
not interfere with the new Interactor's iteration over answers. 
Backtracking over the Interactor's creation point, as such, 
makes it unreachable and therefore subject to garbage collection.

An Interactor is stopped with the 
\begin{codex}
stop(Interactor)
\end{codex}
operation, that might or might not reclaim resources held by the engine.
In our later implementation {\em Lean Prolog}, we are using a fully
automated memory management mechanism where unreachable
engines are automatically garbage collected.
While this API clearly refers to operations
going beyond Horn Clause logic, it can be shown
that a fairly high-level pure Prolog semantics can 
be given to them in a style somewhat similar
to what one would do when writing a Prolog
interpreter in Haskell, as shown in section 4 of
\cite{padl09inter}.

So far, these operations provide a minimal API, powerful
enough to switch tasks cooperatively between an engine and its
``client"\footnote{Another Prolog engine using and engine's services}
and emulate key Prolog built-ins like
{\tt if-then-else} and {\tt findall} \cite{tarau:cl2000},
as well as higher order operations like 
{\em fold} and {\em best\_of} \cite{padl09inter}.
We  give more details on emulations
of these constructs in section \ref{srcext}.

\subsection{A yield/return operation} \label{return}

The following operations provide a ``mixed-initiative'' 
interaction mechanism, allowing more general data exchanges 
between an engine and its client.

First, like the {\tt yield return} construct of \verb~C#~ 
and the {\tt yield operation} of Ruby and Python, 
our {\tt return/1} operation 
\begin{codex}
return(Term)
\end{codex}
\noindent saves the state of the engine, and transfers 
{\em control} and a {\em result} {\tt Term} to its client. The client 
receives a copy of {\tt Term} when using 
its {\tt get/2} operation. 

Note that an Interactor returns control to its client 
either by calling {\tt return/1}
or when a computed answer becomes available.
By using a sequence of {\tt return/get} operations, 
an engine can provide a stream of {\em intermediate/final results} 
to its client, without having to backtrack. This mechanism 
is powerful enough to implement a complete exception 
handling mechanism simply by defining
\begin{codex}
throw(E):-return(exception(E)).
\end{codex}
\noindent When combined with a {\tt catch(Goal,Exception,OnException)}, 
on the client side, the client can decide,
upon reading the exception with {\tt get/2}, if it wants 
to handle it or to throw it to the next level.

\subsection{Coroutining Logic Engines} \label{yield}

Coroutining has been in use in Prolog systems mostly to implement
constraint programming extensions. The typical mechanism
involves {\em attributed variables} holding suspended goals
that may be triggered by changes in the instantiation state
of the variables. We  discuss here a different form of
coroutining, induced by the ability to switch back and forth
between engines.
    
The operations described so far allow an engine to return 
answers from any point in its computation sequence. 
The next step is to enable an engine's 
{\em client}\footnote{Another engine, that uses an engine's services.} to {\em inject} 
new goals (executable data) to an arbitrary inner context 
of another engine. Two new primitives are needed:

\begin{codex}
to_engine(Engine,Data)
\end{codex}
\noindent that is called by the client to send data to an Engine, and
\begin{codex}
from_engine(Data)
\end{codex}
that is called by the engine to receive a client's Data.

\medskip
\noindent A typical use case for the {\em Interactor API} looks as follows:
\begin{enumerate}
\item the {\em client} creates and initializes a new {\em engine}
\item the client triggers a new computation in the {\em engine}, 
parameterized as follows:

\begin{enumerate}
\item the {\em client} passes some data and a new goal to the {\em engine} 
and issues a {\tt get} operation that passes control to it
\item the {\em engine} starts a computation from its initial goal or 
the point where it has been suspended and runs (a copy of) 
the new goal received from its {\em client}
\item the {\em engine} returns (a copy of) the answer, 
then suspends and returns control to its {\em client}
\end{enumerate}
\item the {\em client} interprets the answer and proceeds 
with its next computation step
\item the process is fully reentrant and the {\em client} 
may repeat it from an arbitrary point in its computation
\end{enumerate}

Using a metacall mechanism like {\tt call/1}\footnote{Which, interestingly enough, can itself be emulated
in terms of engine operations \cite{tarau:cl2000} or directly through a source
level transformation \cite{Tarau90:PLILP}).}, one can implement a close equivalent
of Ruby's {\tt yield} statement as follows:
\begin{codex}
ask_engine(Engine,(Answer:-Goal), Result):-
  to_engine(Engine,(Answer:-Goal)),
  get(Engine,Result).

engine_yield(Answer):-
  from_engine((Answer:-Goal)),
  call(Goal),
  return(Answer).
\end{codex}
The predicate {\tt ask\_engine/3} sends a query (possibly built 
at runtime) to an engine, which in turn, executes it and returns 
a result with an {\tt engine\_yield} operation. The query
is typically a goal or a pattern of the form {\tt AnswerPattern:-Goal} in
which case the engine interprets it as a request to instantiate
{\tt AnswerPattern} by executing {\tt Goal} before returning the answer
instance.

As the following example shows, this allows the client 
to use, from outside, the (infinite) recursive loop of an engine 
as a form of {\em updatable persistent state}.
\begin{codex}
sum_loop(S1):-engine_yield(S1=>S2),sum_loop(S2).

inc_test(R1,R2):-new_engine(_,sum_loop(0),E),
   ask_engine(E,(S1=>S2:-S2 is S1+2),R1),
   ask_engine(E,(S1=>S2:-S2 is S1+5),R2).
   
?- inc_test(R1,R2).
R1=the(0=>2), R2=the(2=>7).
\end{codex}

\noindent  Note also that after parameters (the increments 2 and 5)
are passed to the engine, results dependent on its state 
(the sums so far 2 and 7) are received back. Moreover, note 
that an arbitrary goal is injected in the local context of the 
engine where it is executed. The goal can then access the engine's 
{\em state variables} {\tt S1} and {\tt S2}. 
As engines have separate garbage collectors (or in simple 
cases as a result of tail recursion), their infinite loops 
run in constant space, provided that no unbounded size objects 
are created.

\section{Source level extensions through new definitions} \label{srcext}

To give a glimpse of the expressiveness of the resulting Horn Clause + Engines 
language, first described in \cite{tarau:cl2000} we  specify a number of
built-in predicates known as "impossible to emulate" in
Horn Clause Prolog
(except by significantly
lowering the level of abstraction and implementing
something close to the virtual machine itself).

\subsection{Negation, first\_solution/3, if\_then\_else/3}

These constructs are implemented simply by discarding all but
the first solution produced by an engine. The predicate
{\tt first\_solution} (usable to implement {\tt once/1}),
returns {\tt the(X)} or the atom {\tt no}
as first solution of goal {\tt G}:
\begin{code}
first_solution(X,G,Answer):-
  new_engine(X,G,E),
  get(E,R),stop(E),
  Answer=R.

not(G):-first_solution(_,G,no).
\end{code}
The same applies to an emulation of Prolog's if-then-else construct, shown here
as the predicate {\tt if\_then\_else/3}, 
which, if {\tt Cond} succeeds, calls {\tt Then}, keeping the bindings
produced by {\tt Cond} and otherwise calls {\tt Else} after undoing the bindings of the call to 
{\tt Cond}.
\begin{code}
if_then_else(Cond,Then,Else):-
  new_engine(Cond,Cond,E),
  get(E,Answer), stop(E),
  select_then_else(Answer,Cond,Then,Else,Goal),
  Goal.
  
select_then_else(the(Cond),Cond,Then,_Else,Then).  
select_then_else(no,_,_,_Then,Else,Else).  
\end{code}
Note that these operations require the use of {\tt CUT} in typical
Prolog library implementations. While in the presence of
engines, one can control the generation of multiple answers directly
and only use the CUT when more complex control constructs are required
(like in the case of embedded disjunctions),
given the efficient WAM-level implementation of {\tt CUT} and the
frequent use of Prolog's if-then-else construct, emulations of these
built-ins can be seen mostly as an executable
specification of their faster low-level counterparts.

\subsection{Reflective meta-interpreters} \label{reflect}

A simple Horn Clause+Engines meta-interpreter {\bf metacall/1}
just {\em reflects} backtracking through {\bf element\_of/2}
over deterministic engine operations. 

\begin{code}
metacall(Goal):-
  new_engine(Goal,Goal,E),
  element_of(E,Goal).
  
element_of(E,X):-get(E,the(A)),select_from(E,A,X).

select_from(_,A,A).
select_from(E,_,X):-element_of(E,X).
\end{code}

{\noindent We} can see {\tt metacall/1} as an operation which fuses two
orthogonal language features provided by an engine:
{\em computing an answer of a Goal}, and {\em advancing to the next
answer}, through the source level operations {\tt element\_of/2} and
{\tt select\_from/3} which 'borrow' the ability to backtrack from the
underlying interpreter. 
The existence of the simple meta-interpreter defined by {\tt metacall/1} indicates
that first-class engines lift the expressiveness of
Horn Clause logic significantly.

\subsection{All-solution predicates} \label{allsols}

All-solution predicates like {\tt findall/3} can be obtained by collecting
answers through recursion. The (simplified) code
consists of {\tt findall/3} that creates an engine and
{\tt collect\_all\_answers/3} that recurses while
new answers are available.
\begin{code}
findall(X,G,Xs):-
  new_engine(X,G,E), 
  get(E,Answer),
  collect_all_answers(Answer,E,Xs).

collect_all_answers(no,_,[]).
collect_all_answers(the(X),E,[X|Xs]):-get(E,Answer),
  collect_all_answers(Answer,E,Xs).
\end{code}
Note that after the auxiliary engine created for {\tt findall/3}
is discarded, heap space is needed {\em only
to hold the computed answers}, as it is also the case with the conventional
implementation of {\tt findall}. Note also that
the implementation handles embedded uses of {\tt findall} naturally
and that no low-level built-ins are needed.

\subsection{Term copying and instantiation state detection} \label{copyterm}
As standardizing variables in the returned answer is
part of the semantics of {\tt get/2}, term copying
is just computing a first solution to {\tt true/0}.
Implementing {\tt var/1} uses the fact that only free
variables can have copies unifiable with two distinct constants.

\begin{code}
copy_term(X,CX):-first_solution(X,true,the(CX)).

var(X):-copy_term(X,a),copy_term(X,b).
\end{code}

{\noindent The} previous definitions have shown that the resulting
language subsumes (through user provided definitions)
constructs like negation as failure, if-then-else,
once, {\tt copy\_term}, {\tt findall} - this suggests calling
this layer {\em Kernel
Prolog}. As Kernel Prolog
contains negation as failure, following \cite{ISOProlog}
we can, in principle, use it for an executable
specification of full Prolog.

It is important to note here that the engine-based
implementation serves in some cases just as a proof
of expressiveness and that, in practice, operations like
{\tt var/1} for which even a small overhead is unacceptable
are implemented directly as built-ins.
Nevertheless, the engine-based source-level definitions
provide in all cases a reference implementation usable as a specification
for testing purposes.
 
\subsection{Implementing exceptions} \label{exc}

While it is possible to implement an exception mechanism at source level as shown in
\cite{td94:LOPSTR}, through a continuation passing program 
transformation (binarization), one can use engines
for the same purpose.
By returning a new answer pattern as indication of an exception,
a simple and efficient implementation of exceptions is
obtained.

We have actually chosen this implementation scenario
in the BinProlog compiler which also provides a {\bf return/1} operation
to exit an engine's emulator loop with an arbitrary answer pattern,
possibly before the end of a successful derivation. The (somewhat simplified)
code is as follows:
\begin{code}
throw(E):-return(exception(E)).

catch(Goal,Exception,OnException):-
  new_engine(answer(Goal),Goal,Engine),
  element_of(Engine,Answer),
  do_catch(Answer,Goal,Exception,OnException,Engine).
  
do_catch(exception(E),_,Exception,OnException,Engine):-
  (E=Exception->
    OnException % call action if matching
  ; throw(E)    % throw again otherwise
  ), stop(Engine).
do_catch(the(Goal),Goal,_,_,_).
\end{code}

The {\tt throw/1} operation returns a special exception
pattern, while the {\tt catch/3} operation
stops the engine, calls a handler on matching exceptions
or re-throws non-matching ones to the next layer.
If engines are lightweight, the cost of using
them for exception handling is acceptable performance-wise,
most of the time. However, it is also possible to reuse an
engine (using {\tt load\_engine/3}) - 
for instance in an inner loop, to define a handler for
all exceptions that can occur, rather than wrapping up
each call into a new engine with a catch.

\subsection{Interactors and higher order constructs} \label{higher}
As a glimpse at the expressiveness of the Interactor API, we 
implement, in the tradition of higher order functional programming, 
a {\em fold} operation.
The predicate {\tt efoldl} can be seen as a generalization of {\tt findall}
connecting results produced by independent branches 
of a backtracking Prolog engine by applying to them 
a closure {\tt F} using {\tt call/4}:
\begin{codex}
efoldl(Engine,F,R1,R2):-get(Engine,X),efoldl_cont(X,Engine,F,R1,R2).

efoldl_cont(no,_Engine,_F,R,R).
efoldl_cont(the(X),Engine,F,R1,R2):-call(F,R1,X,R),efoldl(Engine,F,R,R2).
\end{codex}

\noindent Classic functional programming idioms like 
{\em reverse as fold} are then implemented simply as: 

\begin{codex} 
reverse(Xs,Ys):-
  new_engine(X,member(X,Xs),E),
  efoldl(E,reverse_cons,[],Ys).  
  
reverse_cons(Y,X,[X|Y]).
\end{codex}
Note also the automatic {\em deforestation} effect 
\cite{journals/tcs/Wadler90} of this programming 
style - no intermediate list structures need to be built, 
if one wants to aggregate the values retrieved from 
an arbitrary generator engine with an operation 
like sum or product.

\section{Extending the Prolog kernel using Interactors} \label{kernelext}

We  review here a few typical extensions of the Prolog kernel 
showing that using first class Logic Engines results in a compact and
portable architecture that is built almost entirely {\em at source level}.

\subsection{Emulating dynamic databases with Interactors} \label{db}

The gain in expressiveness coming directly from the view of 
Logic Engines as iterative answer generators (i.e. Fluents \cite{tarau:cl2000})
is significant. 
The notable exception is Prolog's dynamic database, 
requiring the bidirectional communication provided by interactors.

The key idea for implementing dynamic database operations 
with interactors is to use a logic engine's state in an 
infinite recursive loop.

First, a simple difference-list based infinite server loop is built:
\begin{codex}
queue_server:-queue_server(Xs,Xs).
    
queue_server(Hs1,Ts1):-
  from_engine(Q),server_task(Q,Hs1,Ts1,Hs2,Ts2,A),return(A),
  queue_server(Hs2,Ts2).
\end{codex}

\noindent Next we provide the queue operations, 
needed to maintain the state of the database. To keep the code simple,
we  only focus in this section on operations resulting in additions at
the end of the database.

\begin{codex}
server_task(add_element(X),Xs,[X|Ys],Xs,Ys,yes).
server_task(queue,Xs,Ys,Xs,Ys,Xs-Ys).
server_task(delete_element(X),Xs,Ys,NewXs,Ys,YesNo):-
  server_task_delete(X,Xs,NewXs,YesNo).
\end{codex}

\noindent Then we implement the auxiliary predicates 
supporting various queue operations. 
\begin{codex}
server_task_delete(X,Xs,NewXs,YesNo):-
  select_nonvar(X,Xs,NewXs),!,
  YesNo=yes(X).
server_task_delete(_,Xs,Xs,no).

select_nonvar(X,XXs,Xs):-nonvar(XXs),XXs=[X|Xs].
select_nonvar(X,YXs,[Y|Ys]):-nonvar(YXs),YXs=[Y|Xs],
  select_nonvar(X,Xs,Ys).
\end{codex}
Next, we put it all together, 
as a dynamic database API. 

We can create a 
new engine server providing Prolog
database operations:
\begin{codex}
new_edb(Engine):-new_engine(done,queue_server,Engine).
\end{codex}
We can add new clauses to the database
\begin{codex}
edb_assertz(Engine,Clause):-
  ask_engine(Engine,add_element(Clause),the(yes)).
\end{codex}
and we can return fresh instances of asserted clauses
\begin{codex}  
edb_clause(Engine,Head,Body):-
  ask_engine(Engine,queue,the(Xs-[])),
  member((Head:-Body),Xs).
\end{codex}
or remove them from the the database
\begin{codex}  
edb_retract1(Engine,Head):-Clause=(Head:-_Body),
  ask_engine(Engine,delete_element(Clause),the(yes(Clause))).
\end{codex}
Finally, the database can be discarded by stopping the engine that hosts it:
\begin{codex}  
edb_delete(Engine):-stop(Engine).
\end{codex}

Externally implemented dynamic databases 
can also be made visible as Interactors and reflection 
of the interpreter's own handling of the Prolog 
database becomes possible. 
As an additional benefit, multiple databases can be provided.
This simplifies adding module, object or agent layers at source level. 
By combining database and communication Interactors, 
support for mobile code and autonomous agents can be built 
as shown in \cite{td:tlp}.
Encapsulating external stateful objects like file 
systems, external database or Web service interfaces
as Interactors can provide a uniform interfacing mechanism
and reduce programmer learning curves in Prolog
applications.

A note on practicality is needed here. While indexing can be added
at source level by using hashing on various arguments, the relative
performance compared to compiled code, of this
emulated database is 2-3 orders of magnitude slower. Therefore, in
our various Prolog systems we have used this more as an
executable specification rather than the default implementation
of the database.

\subsection{Refining control: a backtracking if-then-else} \label{ifany}
Various Prolog implementations 
also provide a variant of {\tt if-then-else} 
(called {\tt *->/3} in SWI-Prolog and {\tt if/3} in SICStus-Prolog) 
that either backtracks 
over multiple answers of its guard {\tt Cond} (and calls its {\tt Then} branch for each)
or it switches to the {\tt Else} branch 
if no such answers of {\tt Cond} are found. 
With the same API, we can implement it at source level as follows:

\begin{codex}
if_any(Cond,Then,Else):-
  new_engine(Cond,Cond,Engine),
  get(Engine,Answer),
  select_then_or_else(Answer,Engine,Cond,Then,Else).

select_then_or_else(no,_,_,_,Else):-Else.
select_then_or_else(the(BoundCond),Engine,Cond,Then,_):-
  backtrack_over_then(BoundCond,Engine,Cond,Then).

backtrack_over_then(Cond,_,Cond,Then):-Then.
backtrack_over_then(_,Engine,Cond,Then):-
  get(Engine,the(NewBoundCond)),
  backtrack_over_then(NewBoundCond,Engine,Cond,Then).
\end{codex}

\subsection{Simplifying algorithms: 
Interactors and combinatorial generation} \label{combgen}

Various combinatorial generation algorithms have elegant 
backtracking implementations. However, it is notoriously 
difficult (or inelegant, through the use of ad-hoc side effects) 
to compare answers generated by different OR-branches 
of Prolog's search tree.

\subsubsection{Comparing alternative answers} \label{altanswers}
Optimization problems, selecting the ``best'' 
among answers produced on alternative
branches can easily be expressed as follows:
\begin{itemize}
\item running the generator in a separate logic engine
\item collecting and comparing the answers in a client controlling the engine
\end{itemize}
\noindent The second step can actually be automated, provided that 
the comparison criterion is given as a predicate
\begin{codex}
compare_answers(Comparator,First,Second,Best)
\end{codex}
\noindent to be applied to the engine with an {\tt efold} operation:
\begin{code}
best_of(Answer,Comparator,Generator):-
  new_engine(Answer,Generator,E),
  efoldl(E,compare_answers(Comparator),no,Best),
  Answer=Best.

compare_answers(Comparator,A1,A2,Best):-
  ( A1\==no,call(Comparator,A1,A2)->Best=A1
  ; Best=A2
  ).
\end{code}
\begin{codex}
?-best_of(X,>,member(X,[2,1,4,3])).
X=4
\end{codex}
Note that in the call to {\tt compare\_answers} the closure {\tt compare\_answers(Comparator)},
gets the extra arguments {\tt A1} and {\tt A2} out of which, depending on the comparison, {\tt Best}
is selected at each step of {\tt efoldl}.

\subsubsection{Encapsulating infinite computation streams} \label{infinite}
An infinite stream of natural numbers is implemented as:
\begin{codex}
loop(N):-return(N),N1 is N+1,loop(N1).
\end{codex}

The following example shows a simple space efficient 
generator for the infinite stream of prime numbers:
\begin{code}
prime(P):-prime_engine(E),element_of(E,P).

prime_engine(E):-new_engine(_,new_prime(1),E).

new_prime(N):- N1 is N+1,
  (test_prime(N1) -> true ; return(N1)),
  new_prime(N1).

test_prime(N):-M is integer(sqrt(N)),between(2,M,D),N mod D =:=0
\end{code}
\noindent Note that the program has been wrapped, using 
the {\tt element\_of} predicate 
to provide one answer at a time through backtracking. 
Alternatively, a forward recursing client can use 
the {\tt get(Engine)} operation to extract 
primes one at a time from the stream.

\section{A short history of BinProlog and its derivatives} \label{history}

The first iteration of BinProlog goes back to around 1990.
Along the years it has pioneered some interesting
architectural choices while adopting a number of
new (at the time) implementation ideas from others.
From 1999 on, we have also released a Java port of BinProlog called
Jinni Prolog, using essentially the same runtime system and
compiler as BinProlog and resulting in 
some new developments happening either
on the Java or C side. 
Some of BinProlog's features are interesting to mention mostly
for historical reasons - as they either became
part of various Prolog systems, when genuinely practical,
or, on the contrary, have turned out to have
only limited, program-specific benefits.
Among features for which BinProlog has been either a pioneer or an early adopter in the
world of Prolog implementations, that have not been covered in this paper are:
\begin{itemize}
\item an efficient implementation of {\tt findall} using a heap splitting
  technique resulting in a single copy operation \cite{Tarau92:ECO}  
  \item a multithreading API using native threads  under explicit
        programmer control (around 1992-1993)
  \item a blackboard architecture using Linda coordination between threads
       \cite{pt93a,dbt95a}
  \item backtrackable global variables (around 1993)
  \item a mechanism for ``partial compilation" to C \cite{tdb94,tdb95rev}
  \item using continuations to implement Prolog extensions, including
  catch/throw \cite{td94:LOPSTR}
  \item cyclic terms (originating in Prolog III) and subterm-sharing implemented using
   a space efficient value trailing mechanism (around 1993)
  \item memoing of goal-independent answer substitutions for deterministic
  calls \cite{pt93b,TBD95:memo}
  \item a DCG variant using backtrackable state updates \cite{DT97:AGNL}
  \item on the fly compilation of dynamic code, based on runtime 
        call/update statistics (around
        1994-1995)
        (a technique similar to the  
        {\em HotSpot} compilation now popular in Java VMs) 
  \item segment preserving copying GC \cite{Demoen96:GC}
  \item assumption grammars - a mechanism extending Prolog
  grammar with hypothetical reasoning \cite{DT97:AGNL}
  \item strong mobility of code and data by transporting live continuations between Prolog
  processes \cite{TD98:mobile,td:tlp}
  \item Prolog based shared virtual worlds supporting simple natural language
  interactions \cite{lm}
\end{itemize}
Elements of the BinProlog continuation passing implementation model 
have been successfully reused in a few different
Prolog systems:
\begin{itemize}
\item jProlog (\url{http://www.cs.kuleuven.be/~bmd/PrologInJava/}) 
written in Java, mostly by Bart Demoen with some help 
from Paul Tarau in 1997, used a Prolog to Java translator 
with binarization as a source to source transformation
\item Jinni Prolog, written in Java by 
Paul Tarau (\url{http://www.binnetcorp.com/Jinni}) 
actively developed since 1998, first as continuation 
passing interpreter and later as a BinWAM compiler
\item A Java port of
Free Prolog (a variant
of BinProlog 2.0) by
Peter Wilson \url{http://www.binnetcorp.com/OpenCode/free_prolog.html} (around
1999) with additions and fixes by Paul Tarau
\item Kernel Prolog, a continuation passing interpreter 
written in Java by Paul Tarau 
(\url{http://www.binnetcorp.com/OpenCode/kernelprolog.html}) around 1999
\item PrologCafe (\url{http://kaminari.scitec.kobe-u.ac.jp/PrologCafe/}), 
a fairly complete Prolog system derived from jProlog implemented by  
Mutsunori Banbara and Naoyuki Tamura
\item 
\verb~P#~ derived from PrologCafe, written in \verb~C#~ by 
Jon Cook (\url{http://homepages.inf.ed.ac.uk/jcook/})
\item Carl Friedrich Bolz's Python-based Prolog interpreter and JIT compiler, 
using AND+OR-continuations directly, without a program transformation \cite{bolz2010}
\item Lean Prolog - a new first-class Logic Engines based lightweight Prolog
system using two identical C and Java-based BinWAM runtime systems
to balance performance and flexibility implemented by
Paul Tarau (work in progress, started in 2008)
\end{itemize}

\section{Related work} \label{related}

Most modern Prolog implementations are centered around the 
Warren Abstract Machine (WAM) \cite{WA83,kaci91:WAM}
which has stood amazingly well the test of time.
In this sense BinProlog's BinWAM is no exception,
although its overall "rate of mutations" with respect to the
original WAM is probably comparable to systems
like Neng-Fa Zhou's TOAM or TOAM-Jr \cite{87972,DBLP:conf/iclp/Zhou07}
or Jan Wielemaker's SWI-Prolog \cite{DBLP:conf/iclp/Wielemaker03}
and definitely
higher, if various extensions are factored out, than
the basic architecture of systems like 
GNU-Prolog \cite{DBLP:journals/jflp/DiazC01},
SICStus Prolog \cite{carlsson1263sicstus}, 
Ciao \cite{DBLP:conf/iclp/CarroH99}, 
YAP \cite{DBLP:conf/iclp/SilvaC06} or XSB \cite{IS1994:Swift2}.
We refer to \cite{VanRoy:SequentialPrologImp} and
\cite{728423} for extensive comparisons of compilation techniques
and abstract machines for various logic programming systems.

Techniques for adding built-ins to binary Prolog
are first discussed in \cite{Demoen91:RU}, where
an implementation oriented view of binary programs
that a binary 
program is simply one that does not need an environment in the WAM
is advocated.
Their paper also describes a technique for implementing Prolog's CUT in
a binary Prolog compiler.
Extensions to BinProlog's AND-continuation passing transformation to also cover
OR-continuations are described in \cite{lindgren94}.

Multiple Logic Engines have been present in one form
or another in various parallel implementation of logic programming
languages \cite{72555,DBLP:conf/lp/Ueda85}. Among the earliest examples
of parallel execution mechanisms for Prolog, AND-parallel
\cite{12071} and OR-parallel \cite{Lusk93applicationsof}
execution models are worth mentioning. 

However,
with the exception of this author's
papers on this topic 
\cite{tarau:shaker,tarau:parimp99,tarau:cl2000,td:tlp,iclp08:inter,padl09inter,damp11tarau} 
we have not found 
an extensive use of first-class
Logic Engines as a mechanism to
enhance language expressiveness, independently
of their use for parallel programming,
with maybe the exception of \cite{carro07}
where such an API is discussed for
parallel symbolic languages in general.
In combination with multithreading \cite{damp11tarau},
our own engine-based API bears similarities
with various other Prolog systems,
notably \cite{DBLP:conf/iclp/CarroH99,DBLP:conf/iclp/Wielemaker03}
while focusing on uncoupling ``concurrency for performance" and
``concurrency for expressiveness".

The use of a garbage collected, infinitely looping
recursive program to encapsulate state
goes back to early work in logic programming 
and it is likely to be common in implementing
various server programs. However
an infinitely recursive pure Horn Clause program
is an ``information sink" that does
not communicate with the outside world
on its own. The minimal
API ({\tt to\_engine/2} and {\tt from\_engine/1})
described in this paper provides
interoperation with such programs, in a generic way.

\section{Conclusion} \label{concl}

At the time of writing,
BinProlog has been around for almost 20
years.
As mostly a single-implementor system, 
BinProlog has not kept
up with systems that have benefited from a larger
implementation effort in terms of
optimizations and extensions like
constraints or tabling,
partly also because our research interests
have diverged towards areas as diverse as
natural language processing, logic synthesis
or computational mathematics.
On the other hand,
re-implementations in Java and a number of
experimental features make it still
relevant as a due member of the unusually
rich and colorful family of Prolog systems.


Our own re-implementations of BinProlog's
virtual machine have been extended with
first-class Logic Engines that
can be used to build on top of 
pure Prolog a practical Prolog system, 
including dynamic database operations, 
entirely at source level.
In a broader sense, interactors can be seen as a starting 
point for rethinking fundamental programming language 
constructs like Iterators and Coroutining in terms 
of language constructs inspired by {\em performatives} 
in agent oriented programming. 
Along these lines, we are currently building a new BinWAM based 
implementation, {\em Lean Prolog}, that combines
a minimal WAM kernel with an almost entirely source-level
{\em interactor}-based 
implementation of Prolog's built-ins and libraries.
We believe that under this new incarnation
some of BinProlog's architectural choices 
are likely to 
have an interesting impact
on the design and implementation of 
future logic programming languages.

\section*{Acknowledgements}
We are thankful to Mar\'ia Garc\'ia de la Banda, Bart Demoen, Manuel Hermenegildo,
Joachim Schimpf
and D.S. Warren and to the anonymous reviewers
for their careful reading and thoughtful comments on earlier drafts of this paper.

Special thanks go to Koen De Bosschere, Bart Demoen, Geert Engels, Ulrich
Neumerkel, Satyam Tyagi and Peter Wilson for their contribution to the
implementation of BinProlog and Jinni Prolog components. 

Richard O'Keefe's public
domain Prolog parser and writer have 
been instrumental in turning BinProlog into a
self-contained Prolog system quickly.

Fruitful discussions with Hassan A\"it-Kaci, Patrice Boizumault, Michel Boyer, Mats Carlsson,
Jacques Cohen, Veronica Dahl, Bart Demoen, Thomas Lindgren, Arun Majumdar,  
Olivier Ridoux, Kostis Sagonas, Sten-{\AA}ke T\"arnlund, D.S. Warren, 
Neng-Fa Zhou and comments 
from a large number of BinProlog users helped improving its design 
and implementation.

\bibliographystyle{acmtrans} 

\begin{thebibliography}{}

\bibitem[\protect\citeauthoryear{A{\"\i}t-Kaci}{A{\"\i}t-Kaci}{1991}]{kaci91:W%
AM}
{\sc A{\"\i}t-Kaci, H.} 1991.
\newblock {\em {Warren}'s Abstract Machine: A Tutorial Reconstruction}.
\newblock MIT Press.

\bibitem[\protect\citeauthoryear{Bolz, Leuschel, and Schneider}{Bolz
  et~al\mbox{.}}{2010}]{bolz2010}
{\sc Bolz, C.~F.}, {\sc Leuschel, M.}, {\sc and} {\sc Schneider, D.} 2010.
\newblock {Towards a Jitting VM for Prolog Execution}.
\newblock In {\em PPDP '10: Proceedings of the 12th international ACM SIGPLAN
  symposium on Principles and practice of declarative programming}. ACM, New
  York, NY, USA, 99--108.

\bibitem[\protect\citeauthoryear{Carlsson, Widen, Andersson, Andersson, Boortz,
  Nilsson, and Sjoland}{Carlsson et~al\mbox{.}}{}]{carlsson1263sicstus}
{\sc Carlsson, M.}, {\sc Widen, J.}, {\sc Andersson, J.}, {\sc Andersson, S.},
  {\sc Boortz, K.}, {\sc Nilsson, H.}, {\sc and} {\sc Sjoland, T.}
\newblock {SICStus Prolog user's manual}.

\bibitem[\protect\citeauthoryear{Carro and Hermenegildo}{Carro and
  Hermenegildo}{1999}]{DBLP:conf/iclp/CarroH99}
{\sc Carro, M.} {\sc and} {\sc Hermenegildo, M.~V.} 1999.
\newblock {Concurrency in Prolog Using Threads and a Shared Database}.
\newblock In {\em ICLP}. 320--334.

\bibitem[\protect\citeauthoryear{Casas, Carro, and Hermenegildo}{Casas
  et~al\mbox{.}}{2007}]{carro07}
{\sc Casas, A.}, {\sc Carro, M.}, {\sc and} {\sc Hermenegildo, M.} 2007.
\newblock Towards a high-level implementation of flexible parallelism
  primitives for symbolic languages.
\newblock In {\em PASCO '07: Proceedings of the 2007 international workshop on
  Parallel symbolic computation}. ACM, New York, NY, USA, 93--94.

\bibitem[\protect\citeauthoryear{Clark and Green}{Clark and
  Green}{1977}]{Clark:1977:ESL:359423.359427}
{\sc Clark, D.~W.} {\sc and} {\sc Green, C.~C.} 1977.
\newblock An empirical study of list structure in lisp.
\newblock {\em Commun. ACM\/}~{\em 20}, 78--87.

\bibitem[\protect\citeauthoryear{Conway}{Conway}{1963}]{coroutinesconway1963}
{\sc Conway, M.~E.} 1963.
\newblock {Design of a Separable Transition-Diagram Compiler}.
\newblock {\em Communications of the ACM\/}~{\em 6,\/}~7, 396--408.

\bibitem[\protect\citeauthoryear{da~Silva and Costa}{da~Silva and
  Costa}{2006}]{DBLP:conf/iclp/SilvaC06}
{\sc da~Silva, A.~F.} {\sc and} {\sc Costa, V.~S.} 2006.
\newblock The design and implementation of the yap compiler: An optimizing
  compiler for logic programming languages.
\newblock In {\em ICLP}, {S.~Etalle} {and} {M.~Truszczynski}, Eds. Lecture
  Notes in Computer Science, vol. 4079. Springer, 461--462.

\bibitem[\protect\citeauthoryear{Dahl, Tarau, and Li}{Dahl
  et~al\mbox{.}}{1997}]{DT97:AGNL}
{\sc Dahl, V.}, {\sc Tarau, P.}, {\sc and} {\sc Li, R.} 1997.
\newblock Assumption {G}rammars for {P}rocessing {N}atural {L}anguage.
\newblock In {\em {Proceedings of the Fourteenth International Conference on
  Logic Programming}}, {L.~Naish}, Ed. MIT press, 256--270.

\bibitem[\protect\citeauthoryear{De~Bosschere and Tarau}{De~Bosschere and
  Tarau}{1996}]{dbt95a}
{\sc De~Bosschere, K.} {\sc and} {\sc Tarau, P.} 1996.
\newblock Blackboard-based {E}xtensions in {P}rolog.
\newblock {\em Software --- Practice and Experience\/}~{\em 26,\/}~1 (Jan.),
  49--69.

\bibitem[\protect\citeauthoryear{Demoen}{Demoen}{1992}]{DBLP:conf/lopstr/Demoe%
n92}
{\sc Demoen, B.} 1992.
\newblock On the transformation of a prolog program to a more efficient binary
  program.
\newblock In {\em LOPSTR}. 242--252.

\bibitem[\protect\citeauthoryear{Demoen, Engels, and Tarau}{Demoen
  et~al\mbox{.}}{1996}]{Demoen96:GC}
{\sc Demoen, B.}, {\sc Engels, G.}, {\sc and} {\sc Tarau, P.} 1996.
\newblock Segment {P}reserving {C}opying {G}arbage {C}ollection for {WAM} based
  {P}rolog.
\newblock In {\em Proceedings of the 1996 ACM Symposium on Applied Computing}.
  ACM Press, Philadelphia, 380--386.

\bibitem[\protect\citeauthoryear{Demoen and Mari\"{e}n}{Demoen and
  Mari\"{e}n}{1992}]{Demoen91:RU}
{\sc Demoen, B.} {\sc and} {\sc Mari\"{e}n, A.} 1992.
\newblock {I}mplementation of {P}rolog as binary definite {P}rograms.
\newblock In {\em Logic Programming, RCLP Proceedings}, {A.~Voronkov}, Ed.
  Number 592 in Lecture Notes in Artificial Intelligence. Springer-Verlag,
  Berlin, Heidelberg, 165--176.

\bibitem[\protect\citeauthoryear{Demoen and Nguyen}{Demoen and
  Nguyen}{2000}]{728423}
{\sc Demoen, B.} {\sc and} {\sc Nguyen, P.-L.} 2000.
\newblock So many wam variations, so little time.
\newblock In {\em CL '00: Proceedings of the First International Conference on
  Computational Logic}. Springer-Verlag, London, UK, 1240--1254.

\bibitem[\protect\citeauthoryear{Deransart, Ed-Dbali, and Cervoni}{Deransart
  et~al\mbox{.}}{1996}]{ISOProlog}
{\sc Deransart, P.}, {\sc Ed-Dbali, A.}, {\sc and} {\sc Cervoni, L.} 1996.
\newblock {\em {Prolog: The Standard}}.
\newblock Springer-Verlag, Berlin.
\newblock ISBN: 3-540-59304-7.

\bibitem[\protect\citeauthoryear{Diaz and Codognet}{Diaz and
  Codognet}{2001}]{DBLP:journals/jflp/DiazC01}
{\sc Diaz, D.} {\sc and} {\sc Codognet, P.} 2001.
\newblock Design and implementation of the gnu prolog system.
\newblock {\em Journal of Functional and Logic Programming\/}~{\em 2001,\/}~6.

\bibitem[\protect\citeauthoryear{FIPA}{FIPA}{1997}]{fipa2:97}
{\sc FIPA}. 1997.
\newblock {FIPA} 97 specification part 2: Agent communication language.
\newblock Version 2.0.

\bibitem[\protect\citeauthoryear{Hermenegildo}{Hermenegildo}{1986}]{12071}
{\sc Hermenegildo, M.~V.} 1986.
\newblock An abstract machine for restricted and-parallel execution of logic
  programs.
\newblock In {\em Proceedings on Third international conference on logic
  programming}. Springer-Verlag New York, Inc., New York, NY, USA, 25--39.

\bibitem[\protect\citeauthoryear{Lindgren}{Lindgren}{1994}]{lindgren94}
{\sc Lindgren, T.} 1994.
\newblock A continuation-passing style for prolog.
\newblock In {\em ILPS '94: Proceedings of the 1994 International Symposium on
  Logic programming}. MIT Press, Cambridge, MA, USA, 603--617.

\bibitem[\protect\citeauthoryear{Liskov, Atkinson, Bloom, Moss, Schaffert,
  Scheifler, and Snyder}{Liskov et~al\mbox{.}}{1981}]{DBLP:books/sp/Liskov81}
{\sc Liskov, B.}, {\sc Atkinson, R.~R.}, {\sc Bloom, T.}, {\sc Moss, J. E.~B.},
  {\sc Schaffert, C.}, {\sc Scheifler, R.}, {\sc and} {\sc Snyder, A.} 1981.
\newblock {\em {CLU Reference Manual}}. Lecture Notes in Computer Science, vol.
  114.
\newblock Springer.

\bibitem[\protect\citeauthoryear{Lloyd}{Lloyd}{1987}]{LL87}
{\sc Lloyd, J.} 1987.
\newblock {\em {F}oundations of {L}ogic {P}rogramming}.
\newblock Symbolic computation --- Artificial Intelligence. Springer-Verlag,
  Berlin.
\newblock Second edition.

\bibitem[\protect\citeauthoryear{Lusk, Mudambi, Gmbh, and Overbeek}{Lusk
  et~al\mbox{.}}{1993}]{Lusk93applicationsof}
{\sc Lusk, E.}, {\sc Mudambi, S.}, {\sc Gmbh, E.}, {\sc and} {\sc Overbeek, R.}
  1993.
\newblock Applications of the aurora parallel prolog system to computational
  molecular biology.
\newblock In {\em In Proc. of the JICSLP'92 Post-Conference Joint Workshop on
  Distributed and Parallel Implementations of Logic Programming Systems,
  Washington DC}. MIT Press.

\bibitem[\protect\citeauthoryear{N\"{a}ss\'{e}n, Carlsson, and
  Sagonas}{N\"{a}ss\'{e}n et~al\mbox{.}}{2001}]{Nassen:2001:IMS:773184.773191}
{\sc N\"{a}ss\'{e}n, H.}, {\sc Carlsson, M.}, {\sc and} {\sc Sagonas, K.} 2001.
\newblock Instruction merging and specialization in the sicstus prolog virtual
  machine.
\newblock In {\em Proceedings of the 3rd ACM SIGPLAN international conference
  on Principles and practice of declarative programming}. PPDP '01. ACM, New
  York, NY, USA, 49--60.

\bibitem[\protect\citeauthoryear{Neumerkel}{Neumerkel}{1992}]{Neum92}
{\sc Neumerkel, U.} 1992.
\newblock {Specialization of Prolog Programs with Partially Static Goals and
  Binarization}.
\newblock Phd thesis.
\newblock Technische Universit\"{a}t Wien.

\bibitem[\protect\citeauthoryear{Shapiro}{Shapiro}{1989}]{72555}
{\sc Shapiro, E.} 1989.
\newblock The family of concurrent logic programming languages.
\newblock {\em ACM Comput. Surv.\/}~{\em 21,\/}~3, 413--510.

\bibitem[\protect\citeauthoryear{Swift and Warren}{Swift and
  Warren}{1994}]{IS1994:Swift2}
{\sc Swift, T.} {\sc and} {\sc Warren, D.~S.} 1994.
\newblock An abstract machine for {SLG} resolution: definite programs.
\newblock In {\em Logic Programming - Proceedings of the 1994 International
  Symposium}, {M.~Bruynooghe}, Ed. The MIT Press, Massachusetts Institute of
  Technology, 633--652.

\bibitem[\protect\citeauthoryear{Tarau}{Tarau}{1991}]{Tarau91:JAP}
{\sc Tarau, P.} 1991.
\newblock A {S}implified {A}bstract {M}achine for the {E}xecution of {B}inary
  {M}etaprograms.
\newblock In {\em Proceedings of the Logic Programming Conference'91}. ICOT,
  Tokyo, 119--128.

\bibitem[\protect\citeauthoryear{Tarau}{Tarau}{1992}]{Tarau92:ECO}
{\sc Tarau, P.} 1992.
\newblock {E}cological {M}emory {M}anagement in a {C}ontinuation {P}assing
  {P}rolog {E}ngine.
\newblock In {\em Memory Management International Workshop IWMM 92
  Proceedings}, {Y.~Bekkers} {and} {J.~Cohen}, Eds. Number 637 in Lecture Notes
  in Computer Science. Springer, 344--356.

\bibitem[\protect\citeauthoryear{Tarau}{Tarau}{1998}]{T98:jelia}
{\sc Tarau, P.} 1998.
\newblock {Towards Inference and Computation Mobility: The Jinni Experiment}.
\newblock In {\em {Proceedings of JELIA'98, LNAI 1489}}, {J.~Dix} {and}
  {U.~Furbach}, Eds. Springer, Dagstuhl, Germany, 385--390.
\newblock invited talk.

\bibitem[\protect\citeauthoryear{Tarau}{Tarau}{1999a}]{tarau:shaker}
{\sc Tarau, P.} 1999a.
\newblock {Inference and Computation Mobility with Jinni}.
\newblock In {\em {The Logic Programming Paradigm: a 25 Year Perspective}},
  {K.~Apt}, {V.~Marek}, {and} {M.~Truszczynski}, Eds. Springer, 33--48.
\newblock ISBN 3-540-65463-1.

\bibitem[\protect\citeauthoryear{Tarau}{Tarau}{1999b}]{tarau:paam99}
{\sc Tarau, P.} 1999b.
\newblock {Intelligent Mobile Agent Programming at the Intersection of Java and
  Prolog}.
\newblock In {\em {Proceedings of The Fourth International Conference on The
  Practical Application of Intelligent Agents and Multi-Agents}}. London, U.K.,
  109--123.

\bibitem[\protect\citeauthoryear{Tarau}{Tarau}{1999c}]{tarau:parimp99}
{\sc Tarau, P.} 1999c.
\newblock {Multi-Engine Horn Clause Prolog}.
\newblock In {\em {Proceedings of Workshop on Parallelism and Implementation
  Technology for (Constraint) Logic Programming Languages}}, {G.~Gupta} {and}
  {E.~Pontelli}, Eds. Las Cruces, NM.

\bibitem[\protect\citeauthoryear{Tarau}{Tarau}{2000}]{tarau:cl2000}
{\sc Tarau, P.} 2000.
\newblock {Fluents: A Refactoring of Prolog for Uniform Reflection and
  Interoperation with External Objects}.
\newblock In {\em {Computational Logic--CL 2000: First International
  Conference}}, {J.~Lloyd}, Ed. London, UK.
\newblock LNCS 1861, Springer-Verlag.

\bibitem[\protect\citeauthoryear{Tarau}{Tarau}{2004a}]{iclp04:jinni}
{\sc Tarau, P.} 2004a.
\newblock {Agent Oriented Logic Programming Constructs in Jinni 2004}.
\newblock In {\em {Logic Programming, 20-th International Conference, ICLP
  2004}}, {B.~Demoen} {and} {V.~Lifschitz}, Eds. Springer, LNCS 3132,
  Saint-Malo, France, 477--478.

\bibitem[\protect\citeauthoryear{Tarau}{Tarau}{2004b}]{ciclops:jinni}
{\sc Tarau, P.} 2004b.
\newblock {Orthogonal Language Constructs for Agent Oriented Logic
  Programming}.
\newblock In {\em {Proceedings of CICLOPS 2004, Fourth Colloquium on
  Implementation of Constraint and Logic Programming Systems}}, {M.~Carro}
  {and} {J.~F. Morales}, Eds. Saint-Malo, France.

\bibitem[\protect\citeauthoryear{Tarau}{Tarau}{2006}]{bp7advanced}
{\sc Tarau, P.} 2006.
\newblock {BinProlog 11.x Professional Edition: Advanced BinProlog Programming
  and Extensions Guide}.
\newblock Tech. rep., BinNet Corp.

\bibitem[\protect\citeauthoryear{Tarau}{Tarau}{2008a}]{iclp08:inter}
{\sc Tarau, P.} 2008a.
\newblock { Logic Engines as Interactors}.
\newblock In {\em {Logic Programming, 24-th International Conference, ICLP}},
  {M.~Garcia de~la Banda} {and} {E.~Pontelli}, Eds. Springer, LNCS, Udine,
  Italy, 703--707.

\bibitem[\protect\citeauthoryear{Tarau}{Tarau}{2008b}]{j2k_ug}
{\sc Tarau, P.} 2008b.
\newblock {The Jinni Prolog Compiler: a fast and flexible Prolog-in-Java}.
\newblock http://www.binnetcorp.com/download/jinnidemo/JinniUserGuide.html.

\bibitem[\protect\citeauthoryear{Tarau}{Tarau}{2011}]{damp11tarau}
{\sc Tarau, P.} 2011.
\newblock Concurrent programming constructs in multi-engine prolog.
\newblock In {\em Proceedings of DAMP'11: ACM SIGPLAN Workshop on Declarative
  Aspects of Multicore Programming}. ACM, New York, NY, USA.

\bibitem[\protect\citeauthoryear{Tarau and Boyer}{Tarau and
  Boyer}{1990}]{Tarau90:PLILP}
{\sc Tarau, P.} {\sc and} {\sc Boyer, M.} 1990.
\newblock {E}lementary {L}ogic {P}rograms.
\newblock In {\em Proceedings of Programming Language Implementation and Logic
  Programming}, {P.~Deransart} {and} {J.~Maluszy{\'n}ski}, Eds. Number 456 in
  Lecture Notes in Computer Science. Springer, 159--173.

\bibitem[\protect\citeauthoryear{Tarau and Boyer}{Tarau and
  Boyer}{1993}]{Tarau93:CONS}
{\sc Tarau, P.} {\sc and} {\sc Boyer, M.} 1993.
\newblock Nonstandard {A}nswers of {E}lementary {L}ogic {P}rograms.
\newblock In {\em Constructing Logic Programs}, {J.~Jacquet}, Ed. J.Wiley,
  279--300.

\bibitem[\protect\citeauthoryear{Tarau and Dahl}{Tarau and
  Dahl}{1994}]{td94:LOPSTR}
{\sc Tarau, P.} {\sc and} {\sc Dahl, V.} 1994.
\newblock Logic {P}rogramming and {L}ogic {G}rammars with {F}irst-order
  {C}ontinuations.
\newblock In {\em Proceedings of LOPSTR'94, LNCS, Springer}. Pisa.

\bibitem[\protect\citeauthoryear{Tarau and Dahl}{Tarau and
  Dahl}{1998}]{TD98:mobile}
{\sc Tarau, P.} {\sc and} {\sc Dahl, V.} 1998.
\newblock {Mobile Threads through First Order Continuations}.
\newblock In {\em {Proceedings of APPAI-GULP-PRODE'98}}. Coruna, Spain.

\bibitem[\protect\citeauthoryear{Tarau and Dahl}{Tarau and Dahl}{2001}]{td:tlp}
{\sc Tarau, P.} {\sc and} {\sc Dahl, V.} 2001.
\newblock {High-Level Networking with Mobile Code and First Order
  AND-Continuations}.
\newblock {\em {Theory and Practice of Logic Programming}\/}~{\em 1,\/}~3
  (May), 359--380.
\newblock Cambridge University Press.

\bibitem[\protect\citeauthoryear{Tarau and De~Bosschere}{Tarau and
  De~Bosschere}{1993a}]{pt93a}
{\sc Tarau, P.} {\sc and} {\sc De~Bosschere, K.} 1993a.
\newblock Blackboard {B}ased {L}ogic {P}rogramming in {B}in{P}rolog.
\newblock In {\em Proceedings of the fifth University of New Brunswick
  Artificial Intelligence Symposium}, {L.~Goldfarb}, Ed. Fredericton, N.B.,
  137--147.

\bibitem[\protect\citeauthoryear{Tarau and De~Bosschere}{Tarau and
  De~Bosschere}{1993b}]{pt93b}
{\sc Tarau, P.} {\sc and} {\sc De~Bosschere, K.} 1993b.
\newblock {Memoing with Abstract Answers and Delphi Lemmas}.
\newblock In {\em {Logic Program Synthesis and Transformation}}, {Y.~Deville},
  Ed. Springer-Verlag. Louvain-la-Neuve, 196--209.

\bibitem[\protect\citeauthoryear{Tarau, De~Bosschere, Dahl, and
  Rochefort}{Tarau et~al\mbox{.}}{1999}]{lm}
{\sc Tarau, P.}, {\sc De~Bosschere, K.}, {\sc Dahl, V.}, {\sc and} {\sc
  Rochefort, S.} 1999.
\newblock {LogiMOO: an Extensible Multi-User Virtual World with Natural
  Language Control}.
\newblock {\em Journal of Logic Programming\/}~{\em 38,\/}~3 (Mar.), 331--353.

\bibitem[\protect\citeauthoryear{Tarau, De~Bosschere, and Demoen}{Tarau
  et~al\mbox{.}}{1996}]{tdb95rev}
{\sc Tarau, P.}, {\sc De~Bosschere, K.}, {\sc and} {\sc Demoen, B.} 1996.
\newblock Partial {T}ranslation: Towards a {P}ortable and {E}fficient {P}rolog
  {I}mplementation {T}echnology.
\newblock {\em Journal of Logic Programming\/}~{\em 29,\/}~1--3 (Nov.), 65--83.

\bibitem[\protect\citeauthoryear{Tarau, De~Bosschere, and Demoen}{Tarau
  et~al\mbox{.}}{1997}]{TBD95:memo}
{\sc Tarau, P.}, {\sc De~Bosschere, K.}, {\sc and} {\sc Demoen, B.} 1997.
\newblock On {D}elphi {L}emmas {A}nd other {M}emoing {T}echniques {F}or
  {D}eterministic {L}ogic {P}rograms.
\newblock {\em Journal of Logic Programming\/}~{\em 30,\/}~2 (Feb.), 145--163.

\bibitem[\protect\citeauthoryear{Tarau, Demoen, and De~Bosschere}{Tarau
  et~al\mbox{.}}{1994}]{tdb94}
{\sc Tarau, P.}, {\sc Demoen, B.}, {\sc and} {\sc De~Bosschere, K.} 1994.
\newblock The {P}ower of {P}artial {T}ranslation: an {E}xperiment with the
  {C}-ification of {B}inary {P}rolog.
\newblock In {\em Proceedings of the First COMPULOG-NOE Area Meeting on
  Parallelism and Implementation Technology}, {M.~Garc\'\i a de~la Banda, J.}
  {and} {H.~M.}, Eds. Madrid/Spain, 3--17.

\bibitem[\protect\citeauthoryear{Tarau and Majumdar}{Tarau and
  Majumdar}{2009}]{padl09inter}
{\sc Tarau, P.} {\sc and} {\sc Majumdar, A.} 2009.
\newblock {Interoperating Logic Engines}.
\newblock In {\em {Practical Aspects of Declarative Languages, 11th
  International Symposium, PADL 2009}}. Springer, LNCS 5418, Savannah, Georgia,
  137--151.

\bibitem[\protect\citeauthoryear{Tarau and Neumerkel}{Tarau and
  Neumerkel}{1993}]{Tarau93:comp}
{\sc Tarau, P.} {\sc and} {\sc Neumerkel, U.} 1993.
\newblock Compact {R}epresentation of {T}erms and {I}nstructions in the
  {B}in{WAM}.
\newblock Tech. Rep. 93-3, Dept. d'Informatique, Universit\'{e} de Moncton.
  Nov.
\newblock available by ftp from clement.info.umoncton.ca.

\bibitem[\protect\citeauthoryear{Tyagi and Tarau}{Tyagi and
  Tarau}{2001}]{padl_java}
{\sc Tyagi, S.} {\sc and} {\sc Tarau, P.} 2001.
\newblock {A Most Specific Method Finding Algorithm for Reflection Based
  Dynamic Prolog-to-Java Interfaces}.
\newblock In {\em {Proceedings of PADL'2001}}, {I.~Ramakrishan} {and}
  {G.~Gupta}, Eds. Las Vegas.
\newblock Springer-Verlag.

\bibitem[\protect\citeauthoryear{Ueda}{Ueda}{1985}]{DBLP:conf/lp/Ueda85}
{\sc Ueda, K.} 1985.
\newblock Guarded horn clauses.
\newblock In {\em Logic Programming '85, Proceedings of the 4th Conference,
  Tokyo, Japan, July 1-3, 1985}, {E.~Wada}, Ed. Lecture Notes in Computer
  Science, vol. 221. Springer, 168--179.

\bibitem[\protect\citeauthoryear{Van~Roy}{Van~Roy}{1994}]{VanRoy:SequentialPro%
logImp}
{\sc Van~Roy, P.} 1994.
\newblock 1983-1993: The wonder years of sequential prolog implementation.
\newblock {\em Journal of Logic Programming\/}.

\bibitem[\protect\citeauthoryear{Wadler}{Wadler}{1990}]{journals/tcs/Wadler90}
{\sc Wadler, P.} 1990.
\newblock Deforestation: Transforming programs to eliminate trees.
\newblock {\em Theor. Comput. Sci.\/}~{\em 73,\/}~2, 231--248.

\bibitem[\protect\citeauthoryear{Warren}{Warren}{1983}]{WA83}
{\sc Warren, D. H.~D.} 1983.
\newblock An {A}bstract {P}rolog {I}nstruction {S}et.
\newblock Technical Note 309, SRI International. Oct.

\bibitem[\protect\citeauthoryear{Wielemaker}{Wielemaker}{2003}]{DBLP:conf/iclp%
/Wielemaker03}
{\sc Wielemaker, J.} 2003.
\newblock {Native Preemptive Threads in SWI-Prolog}.
\newblock In {\em ICLP}, {C.~Palamidessi}, Ed. Lecture Notes in Computer
  Science, vol. 2916. Springer, 331--345.

\bibitem[\protect\citeauthoryear{Zhou}{Zhou}{2007}]{DBLP:conf/iclp/Zhou07}
{\sc Zhou, N.-F.} 2007.
\newblock A register-free abstract prolog machine with jumbo instructions.
\newblock In {\em ICLP}, {V.~Dahl} {and} {I.~Niemel{\"a}}, Eds. Lecture Notes
  in Computer Science, vol. 4670. Springer, 455--457.

\bibitem[\protect\citeauthoryear{Zhou, Takagi, and Ushijima}{Zhou
  et~al\mbox{.}}{1990}]{87972}
{\sc Zhou, N.-F.}, {\sc Takagi, T.}, {\sc and} {\sc Ushijima, K.} 1990.
\newblock A matching tree oriented abstract machine for prolog.
\newblock 159--173.

\end{thebibliography}

\end{document}